\documentclass[nonacm=true,sigconf]{acmart}

\copyrightyear{2021}
\acmYear{2021}
\acmConference[CACM]{CACM}
\acmBooktitle{}
\acmPrice{}
\acmDOI{10.1145/3366423.3380302}
\acmISBN{978-1-4503-7023-3/20/04}

\usepackage{xspace}

\newcommand{\fsl}{\textsl}
\newcommand{\fbf}{\textbf}

\begin{document}

\title{Moral and Social Ramifications of Autonomous Vehicles}

\subtitle{A Qualitative Study of the Perceptions of Professional and Semi-Professional Drivers}

\author[Dubljevi{\'c}]{Veljko Dubljevi{\'c}}
\affiliation{%
  \institution{North Carolina State University}
  \city{Raleigh}
  \state{NC}
  \country{United States}
  \postcode{27695}
}
\email{veljko_dubljevic@ncsu.edu}

\author[Douglas]{Sean Douglas}
\affiliation{%
  \institution{North Carolina State University}
  \city{Raleigh}
  \state{NC}
  \country{United States}
  \postcode{27695}
}
\email{sdougla5@ncsu.edu}

\author[Milojevich]{Jovan Milojevich}
\affiliation{%
  \institution{Oklahoma State University}
  \city{Stillwater}
  \state{Oklahoma}
  \country{United States}
  \postcode{74078}
}
\email{jovan.milojevich@okstate.edu}

\author[Ajmeri]{Nirav Ajmeri}
\orcid{0000-0003-3627-097X}
\affiliation{%
  \institution{University of Bristol}
  \city{Bristol}
  \country{United Kingdom}
  \postcode{}
}
\email{nirav.ajmeri@bristol.ac.uk}

\author[Bauer]{William A.~Bauer}
\affiliation{%
  \institution{North Carolina State University}
  \city{Raleigh}
  \state{NC}
  \country{United States}
  \postcode{27695}
}
\email{wabauer@ncsu.edu}

\author[List]{George F.~List}
\affiliation{%
  \institution{North Carolina State University}
  \city{Raleigh}
  \state{NC}
  \country{United States}
  \postcode{27695}
}
\email{gflist@ncsu.edu}

\author[Singh]{Munindar P.~Singh}
\orcid{0000-0003-3599-3893}
\affiliation{%
  \institution{North Carolina State University}
  \city{Raleigh}
  \state{NC}
  \country{United States}
  \postcode{27695}
}
\email{mpsingh@ncsu.edu}

\date{}

\begin{abstract}
Autonomous Vehicles (AVs) raise important social and ethical concerns, especially about accountability, dignity, and justice.  We focus on the specific concerns arising from how AV technology will affect the lives and livelihoods of professional and semi-professional drivers.  Whereas previous studies of such concerns have focused on the opinions of experts, we seek to understand these ethical and societal challenges from the perspectives of the drivers themselves.

To this end, we adopted a qualitative research methodology based on semi-structured interviews.  This is an established social science methodology that helps understand the core concerns of stakeholders in depth by avoiding the biases of superficial methods such as surveys.

We find that whereas drivers agree with the experts that AVs will significantly impact transportation systems, they are apprehensive about the prospects for their livelihoods and dismiss the suggestions that driving jobs are unsatisfying and their profession does not merit protection.

By showing how drivers differ from the experts, our study has ramifications beyond AVs to AI and other advanced technologies.  Our findings suggest that qualitative research applied to the relevant, especially disempowered, stakeholders is essential to ensuring that new technologies are introduced ethically.
\end{abstract}

\settopmatter{printfolios=true}
\maketitle

\section{Introduction}
\label{sec:Introduction}

In \emph{The Outline of History}, H.~G.~\citet[p.~1100]{HG-Wells-21:history} declared that ``Human history becomes more and more a race between education and catastrophe.''  Although Wells wasn't talking about Autonomous Vehicle (AV) technology, he was talking about economic disasters, and AVs are a prime example of a phenomenon with significant risks to society.

AVs have much to offer but they also bring forth potential catastrophes by affecting the lives, livelihoods, and wellbeing of humans.  We focus on the important ethical concerns of human autonomy and dignity, justice, and equity \citep{EGE-18:ethics} concerning AVs.

We posit that since AV deployments involve people and machines, they should be viewed as sociotechnical systems \citep{AIES-18:ethics}.  Previous work on sociotechnical systems in AI pursues challenges in representation and reasoning as a way to provide a computational basis for user requirements.  In contrast, here we tackle the problem of understanding stakeholder perceptions that underlie any such requirements.  Our approach thus exemplifies ``social-systems analysis'' as an interdisciplinary assessment of the human impact of AI that ``thinks through all the possible effects of AI systems on all parties'' \citep[p.~313]{Crawford+Calo-16:AI-blind-spot}.

Previous works largely consider the opinions of ``experts'' (whether technologists, decision makers, or ethicists).  However, we take the stance that to develop a suitable foundation for an ethical response, we must understand the concerns of stakeholders most likely affected by the introduction of AVs.

Accordingly, we study the perceptions and concerns of professional and semi-professional drivers recognizing them as a source of ground knowledge about potential harms and benefits of AVs.  We aim to turn the tables by using their perceptions and concerns to advise the experts.  This process respects the reasonable pluralism in democratic societies in which ``every one of us has different life experiences that inform our values'' \cite[p.~35]{Himmelreich-20:ethics-technology}.

To elicit and understand the perceptions and values of stakeholders requires in-depth conversations in which they are able to articulate and explain their positions.  To this end, we applied a qualitative research methodology based on semi-structured interviews with drivers.  The resulting understanding brings out the ethical ramifications of deploying AVs as perceived by drivers, and can be a useful basis for decision makers formulating policies that will govern AV deployments.

\paragraph{Organization}
The remainder of the paper proceeds as follows.
Section~\ref{sec:Justice} explains how we approach the social
ramifications of AVs.  Section~\ref{sec:Study} explains the motivation and design of our study.
Section~\ref{sec:Results} presents and analyzes the results of our study.
Section~\ref{sec:Conclusions}
presents concluding remarks.

\section{Justice and Social Ramifications}
\label{sec:Justice}

We distinguish two levels of ethical concern in relation to AI and other technologies: micro-and macro-ethics, reflecting a distinction made by \citet{AIES-18:ethics}.  Micro-ethics is concerned with morally charged interactions between relatively small numbers of individuals, as represented by classic ethical theories such as deontology \citep{Kant-1785:morals}, utilitarianism \citep{Mill-1863:utilitarianism}, and virtue theory \citep{Aristotle-350BCE:ethics}.  Macro-ethics is concerned with the ethics of systems.  Through there may be no sharp cutoff between micro-ethics and macro-ethics, clear examples of each exist.

Much of AI research focuses on micro-ethics.  The classic trolley problem scenario \citep{Foot-67:trolley} in research on AVs exemplifies what \citet{Dubljevic+Bauer-20:AVs-society} call ``ethics on the road.''  \citet{Bergmann-18:AV-socio-political} used the trolley problem in a virtual reality study concerned with how people react in a variety of vehicular dilemmas while balancing concern for their own safety against the safety of others.

Macro-ethics focuses on difficult questions like distributive justice \citep{Rawls-99:justice,Sen-09:justice}: How should we structure the basic institutions of society, as well as distribute goods and opportunities?  What kinds of economic incentives will improve society?

Focusing on the macro issues---the socioeconomic impact of AVs---we posit that fairness requires decision makers and relevant leaders to take account of stakeholder perspectives in designing policies to mitigate the negative impacts and enhance the positive impacts of AV technology.

\citet{Rawls-99:justice} presents an influential theoretical model for analyzing the fairness of socioeconomic institutions and outcomes.  His ``original position'' thought experiment provides an objective and illuminating way to think about the basic structure of society and what members of societies owe each other.  In the original position---conceptualized as a hypothetical device, not an actual historical condition or ``state of nature''---people are behind a ``veil of ignorance'' about their individual characteristics; thus, they do not know who they are.

\emph{Reflective equilibrium} \citep{Rawls-99:justice} is a prominent method to test the normative and empirical adequacy of ethical principles.  It implies a coherence procedure of applying general principles from theories to specific moral intuitions or considered moral judgments of stakeholders.  Specifically, self-interested interlocutors propose principles of justice and test them against considered judgments on particular examples.

We see two reasons to adopt a Rawlsian perspective.  First, ``justice as fairness'' highlights the value of equity when talking about the basic structure of society and, we presume, the public shares a significant interest in creating fair results for drivers as AVs become more integrated into our transportation systems.  Second, the original position involves all relevant stakeholders so that policies are chosen fairly.  Specifically, in the later stages of the original position, stakeholders would refine the abstract principles to concrete policies based on relevant new sociopolitical data.

\section{Study Design and Methodology}
\label{sec:Study}

In developing policies concerning the deployment of AVs, decision makers may accept that drivers need protection from emerging technological advances or that they are part of the problem to be solved to successfully deploy AV technology.  Either way, public policies should be informed by the experiences and worldviews of drivers since they would be affected most directly by the deployment of AVs.

\subsection{Method}
\label{sec:Qualitative}

Qualitative research is a social science methodology that collects non-numeric, value-laden data.  Although qualitative research is not generally known in computer science, outside of a few specialties, it is a useful approach for addressing the ethical and social implications of technology.

In this study, we conducted semi-structured one-on-one interviews.  This method is well suited to exploratory studies and helps gather rich and meaningful data that (1) future research can build on, and (2) could provide decision makers with a broader understanding of the phenomena under investigation as they make decisions that will impact the stakeholders \citep{Given-08:qualitative}.

This method enables interviewers to maintain focus on key research questions while facilitating a free exchange between the interviewer and interviewee.  Interviewers followed a script to guide the conversation but varied the conversation as needed.  Specifically, this approach enables the interviewer to ask relevant follow-up questions and the interviewee feels free to expound on anything they feel is important for the interviewer to know, producing as much information as possible from the drivers within a restricted time.

Each interview was transcribed.  Data analysis was carried out concurrently with data collection to enable the integration of information from each step of the process.  Codebook development and coding of responses followed an iterative process.  We developed codes via abductive analysis, a form of qualitative content analysis that combines elements of both induction and deduction \citep{Timmermans+Tavory-12:qualitative}.

Our strategy employed five analytic stages, namely: (1) familiarization with the data through reading of interview transcripts; (2) identification of a thematic framework that reflects the ideas discussed; (3) indexing the data, i.e., identifying patterns across the transcripts; (4) charting the data, i.e., comparing data across identified patterns; and (5) mapping and interpreting the data, i.e., making sense of the data holistically \citep{Ritchie+Spencer-02:qualitative}.

\subsection{Drivers as Stakeholders}
\label{sec:Drivers}

By engaging with affected individuals, we hope to ensure that the voices of this important group of stakeholders are heard during agenda setting as well as during research on AV deployment, consistent with the expectations of social justice.  Recognizing and probing how stakeholders are affected is an important component of any ``power analysis'' \citep[p.~19]{Sandler-14:technology-ethics} of technological changes in society (i.e., recognizing who is empowered and disempowered by changes).

Accordingly, we interviewed drivers who had at least two years of paid driving experience.  We report on nine interviews with professional truck drivers and semi-professional Uber or Lyft drivers.  Our goal was to identify and analyze perceptions among these stakeholders concerning relevant challenges of deploying AVs.

\subsection{Implications of AVs Investigated}
\label{sec:Claims}

Previous studies on this topic are sparse and largely focused on experts.  For instance, \citet{Pettigrew+Fritschi+Norman-18:AV-implications} investigated societal implications of AVs by conducting interviews of expert stakeholders from government at multiple levels, trade unions, law, technology firms, AV manufacturing, academia, and other areas.  Thus, although they interviewed stakeholders in the AV deployment process, those stakeholders did not represent professional and semi-professional drivers, whose livelihoods could be radically affected by AVs.


In contrast, we investigated potential implications of AVs on this key stakeholder and society at large \emph{through the eyes of the driver community}.  We focused on the claims shown in Table~\ref{tab:claims}.  Though these claims are based on previous findings \citep[pp.~4--7]{Pettigrew+Fritschi+Norman-18:AV-implications}, we approach them from the ground-level perspective of drivers.

\begin{table}[htb]
\centering
\caption{Claims about social aspects of AVs that we evaluate.}
\label{tab:claims}
\begin{tabular}{p{0.21\columnwidth} p{0.7\columnwidth}}
\toprule

\fsl{Transportation industry} & AVs' significant impact on transportation is inevitable\\

\fsl{Drivers' expectations} & Employers should be straightforward about changes and options toward the potentially affected stakeholders\\

\fsl{Reskilling} & Reskilling is manageable due to the decade-plus lead time (this, we note, is generous as technological changes sometimes occur more quickly than expected)\\

\fsl{Responsibility to respond} & Responsibility for dealing with these issues falls across all elements of society (government, business, drivers)\\

\fsl{Driving as a profession} & To the extent that driving jobs are ``unsatisfying and potentially unhealthy'' their elimination could be a positive development so long as other opportunities are available\\
\bottomrule
\end{tabular}
\end{table}

\subsection{Study Logistics}

Approval was obtained from the Institutional Review Board of the authors' institution (IRB approval no.\ 20276) before interviews were conducted.  No personally identifying information was collected during the interviews.  Participants were recruited via study flyers.  The interviewees verbally indicated their willingness to participate at the start of audio recording.  Each interviewee received a \$60 gift card as compensation for their time.

\begin{table*}[htb]
\centering
\caption{Interview questions.}
\label{tab:S1-questions}
\begin{tabular}{l l p{0.75\textwidth}}\toprule
\textbf{Category} & Q\# & Question or prompt\\\midrule

Situation & Q1 & ``Please reflect on morally problematic situations concerning driving decisions from your own experience.''\\

 & Q2 & ``Was there an occasion where your split-second decision prevented a crash?''\\

 & Q3 & ``Could you describe what exactly happened and what could have happened?''\\\midrule

Self-driving & Q4 & ``What is your opinion of self-driving technology?''\\

 & Q5 & ``Is it fair?''\\

 & Q6 & ``How soon do you think it (self-driving technology) might be implemented?''\\

 & Q7 & ``Do you have plans for your career if that turns out to be sooner than anticipated?''\\\midrule

Professional driving & Q8 & ``Do you like your current job?''\\

 & Q9 & ``Do you feel meaningfully connected with others by performing your job duties?''\\

 & Q10 & ``Is your job overall a positive or negative experience for you?''\\
 \bottomrule
\end{tabular}
\end{table*}

In total, nine interviews were conducted: four with professional truck drivers (identified as TR in the quotes below), one female and three males, and five with semi-professional Uber or Lyft drivers (all male, identified as UL in the quotes below).  The TR drivers were professional, typically older than the UL drivers, who were mostly students supplementing their income by driving (hence, ``semi-professional'').  The interviews were conducted from December 2019 through March 2020 by the corresponding author or by undergraduate research assistants trained by the corresponding author.  The undergraduate research assistants received an orientation to the study protocol, as well as training in ethics, qualitative research methods, and interview administration.

The number of interviews we conducted is sufficient, as six interviews is generally recognized in the qualitative research literature as leading to 80\% of thematic saturation \citep{Guest+06:how-many-interviews,Kuzel-99:sampling-qualitative}.  Interviews varied in length from 30 to 70 minutes and were all digitally recorded.  We transcribed them and applied the method detailed in Section~\ref{sec:Qualitative}.

The questions asked and topics discussed are shown in Table~\ref{tab:S1-questions}.  Since professional driving was described by experts (not drivers) as ``unsatisfying and potentially unhealthy'' \citep[p.~5]{Pettigrew+Fritschi+Norman-18:AV-implications} we also asked drivers to elaborate on their feelings about the job, whether they felt meaningfully connected to others socially, and whether they perceived professional driving is overall a positive or negative experience.

\section{Results and Discussion}
\label{sec:Results}

\begin{table*}[htb]
\centering
\caption{Comments on inevitability and fairness of automation.}
\label{tab:S2-fairness}

\begin{tabular}{l p{0.80\textwidth}}\toprule
\textbf{Respondent ID} & \textbf{Comments}\\\midrule
UL-1/Allen & ``I'm in a position where I don't rely on this income.  I use it to supplement things.''\newline
It [Uber driving] is a great college job, but not a good career.  It is a very, very sad career that often puts drivers and passengers in danger.''\newline
``As someone that is relatively comfortable with defending themselves, can drive, work late hours, and is in college it is a great job.  But if you are out of college, in your 30's, and you do this for a full-time job, it would be an unimaginable hell.''\\

UL-2/Bob &  ``can't wait for cars to be fully autonomous''\newline
``it would definitely wipe out a lot of jobs but also\ldots we need to look toward the future\ldots and start seeing what industries are going to be created''\newline
``I definitely have plans outside of driving if it becomes obsolete sooner than expected.''\\

UL-3/Charley & ``a lot of people would be put out of a job because of that [AVs]\ldots but that's also\ldots a sad result of capitalism in my opinion.''\newline
``I've always used Lyft as a supplement to my income\ldots so I'm not too worried with that\ldots but I'm sure there will be a lot of people out there who will be negatively affected by it [AVs].''\\

UL-4/Dan & ``AVs are lazy, they would make us dependent and not use our whole brain.''  \\

UL-5/Earl & ``AVs are not fair from the standpoint of Lyft, Uber, Truck, Taxicab drivers, or food delivery drivers\ldots because that's income for them\ldots there are consequences for that degree of automation.''\\

TR-1/Anthony & ``it's going to happen to one degree or another [AVs] but I don't know as if the driver will ever fully be replaced''\newline
``the smaller companies, with a few exceptions, will still rely on traditionally driven vehicles\ldots the first generation [of AVs] will start going to the resale market'' \\

TR-2/Bill & ``To a lot of people, it's not gonna be fair, but to those who stay on top of the things they are going on\ldots it shuts a lot of doors, but it opens a lot more.''\newline
``I think we'll lose a lot of people\ldots and yeah, they gonna hurt''\newline
``I don't expect it to impact my career at all''\\

TR-3/Christina & ``it [AVs] would put a lot of drivers out of jobs\ldots I don't think it's right.''\newline
``I think it's very important to be able to connect to all of them [the people in warehouses] instead of the machine doing all of it.''\\

TR-4/David & ``It [AVs] would suck to take a lot of drivers off the road.''\\
\bottomrule
\end{tabular}
\end{table*}

Our results include the drivers' perceptions about their jobs as well as how AVs might impact their livelihoods and society at large.  Overall, the drivers expressed ethical concerns about the effects AVs could have on society, specifically on non-professional drivers and passengers.  The UL drivers lamented automation's weakness in human interaction (see Table~\ref{tab:S2-fairness}).  Frequently mentioned were \emph{conversation}, \emph{human intuition}, and \emph{human knowledge} as benefits that society would lose with AVs.  The TR drivers were generally more concerned about the \emph{welfare} of the truck-driving community and the \emph{potential harm} AVs could cause to their livelihoods.  In addition, the TR drivers expressed concerns about the \emph{decision-making process of AVs}.  Nevertheless, there was some variation in TR driver responses.  For instance, some stated that automation could be integrated into the industry with drivers learning to adapt to the new technology, whereas others stated that automation offered too many risks---to both drivers and society---and that, as a result, it would be morally wrong to introduce further automation (see TR-3).

Some researchers have asserted that driving jobs are unsatisfactory, boring and unhealthy \citep{Pettigrew+Fritschi+Norman-18:AV-implications,Viscelli-18:driverless}, but our results paint a picture of TR respondents who see the ``joy of driving'' and at the very least, UL drivers who value the supplemental income.  Altogether, the respondents viewed their job as a net positive for themselves and society, and many were against full automation---some were even opposed to partial automation (see TR-3).  Many of the respondents realized that automation was likely coming.  However, almost all thought it was at least 10 years away.  None had given serious thought to alternative careers, should automation come sooner than they anticipated.

\begin{table*}[htb]
\centering
\caption{Comments on effects on the community and society at large.}
\label{tab:S3-effects}

\begin{tabular}{l p{0.80\textwidth}}\toprule
\textbf{Respondent ID} & \textbf{Comments}\\\midrule

UL-1/Allen & ``self-driving\ldots cars\ldots[are] not a solution to transportation issues.  What \emph{is} a solution is public transportation.''\newline
``the gig-economy is just a band-aid on the lack of a social net in this country''\newline
``when you have an autonomous vehicle, you gain some efficiency improvements\ldots but you lose the human factor\ldots like chatting with the driver about lunch recommendations\ldots or these specific things that warrant a tip''\newline
``If you gave people that infrastructure with the improvements of efficiency with autonomous buses, I think more people would use it and we'd all be better off.''\\

UL-2/Bob & ``it [AVs] could eliminate the need for stoplights because everyone would know what they were doing''\newline
``industries are going to be created that way, we can create new jobs with technology''\newline
``generally fair, you just have to determine which outcome causes the least amount of total harm''\newline
``self-driving technologies are a type of philanthropy it almost seems''\\

UL-3/Charley & ``it [AVs] would make a much wider impact on the economy than a lot of people realize''\newline
``I could see how it could be unfair, but that's also capitalism''\\

UL-4/Dan & ``we're going to have a whole bunch of people on the road that really don't even know how to drive''\newline
``There are so many people who utilize Uber as their therapist''\\

UL-5/ Earl &  ``self-driving\ldots needs to be looked at more than the excitement of just pulling off an incredible technology.  And how it impacts people and society, before you start jumping on bandwagons.''  \\

TR-1/Anthony & ``in the urban environment there's plenty of things, whether it's a pedestrian that is walking or something that's just stopped in the street\ldots what does it do when it senses [this]''\\

TR-2/Bill & NA\\

TR-3/Christina & ``Who's going to be accountable for all the mistakes that they [AVs] will do?\ldots Who is going to be responsible [if an accident occurs?]''\newline
``I would not let my kids ride on a school bus without a physical driver\ldots[which] they tried''\newline
``automation at every level will create many many problems that we are not equipped to deal with''\\

TR-4/David & NA\\
\bottomrule
\end{tabular}
\end{table*}

Table~\ref{tab:S2-fairness} is a selection of personal or group-related ethical concerns stated by each respondent, which can be read as concerns regarding automation or AVs.  Notable examples are \emph{loss of jobs} and \emph{loss of safety} for passengers, as the driver is in a position to intervene or call the authorities if one passenger seems to be threatened by another passenger.  This contrasts with Table~\ref{tab:S3-effects}, where general-level concerns from each respondent were listed specifically in regard to society or some larger community.  This was done to see not just to what extent these issues were a concern, but also to express the range of concerns.

\subsection{Micro-Ethics: Safety For or From Others}
\label{sec:Micro}

\begin{table*}[htb]
\centering
\caption{Comments on initial problems and ``bugs'' with AV technology.}
\label{tab:S4-problems}

\begin{tabular}{l p{0.80\textwidth}}\toprule
\textbf{Respondent ID} & \textbf{Comments}\\\midrule

UL-1/Allen & ``others on the road may not understand that your fancy new Tesla may not have all of the bugs worked out''\\

UL-2/Bob & ``there will be growing pains with that, but that's all of technology for all of human history has growing pains.  But that's not something that scares me''\\

UL-3/Charley & ``I think there's just so many factors [for AVs] on the road that I wouldn't want to be on the front end on it--because there's going to be a lot of mistakes made\ldots it's that first phase where they bring it in, I think it's a phase they're trying to get around''\newline
``they don't know what the bugs are in the system until they happen'' \\

UL-4/Dan & ``what happens when someone comes from the side of you and your car doesn't know?  What happens if your car malfunctioned?  \ldots there is no way you're going to tell me that my car can drive itself and still be alert like I am''\newline
``Why weren't your feet on the floor?  Why are you not paying attention?  I think it's [AV] just a crutch''\newline
``what happens when shit malfunctions and things gotta get real?  And you don't have anybody to blame or anything?''  \\

UL-5/ Earl & ``I would be more inclined to wait until they've got more time working out the bugs, I'm weary of it [AV].''\newline
``a lot of that would have to be debugged or tried/experimented with.''\newline
``[Self-driving tech] needs to be looked at more than the excitement of just pulling off an incredible technology''\\

TR-1/Anthony & ``I think the technology can all sense all that stuff, but it's just a matter of what does it so when it senses it?  Does it shut down completely\ldots It's just a lot of testing that I think has to take place before it's widely accepted''\\

TR-2/Bill & NONE\\

TR-3/Christina & ``What if the truck in front of me blows out a tire and it's all kinds of debris on the road?  What is the computer going to know to do in that situation\ldots?''\newline
``there was an incident where they [an AV] killed someone who was crossing the street''\newline
``A Tesla was in an accident and the car completely shut down; you couldn't open the window.  He [the driver] burned alive in the car''\\

TR-4/David & NONE\\
\bottomrule
\end{tabular}
\end{table*}

Note that Table~\ref{tab:S4-problems} is a gauge for the level of awareness of each respondent, not a qualitative difference in definition of problems with AV technology.
Concerning micro-ethics, and specifically safety concerns, we present summaries of responses in Table~\ref{tab:S4-problems}, citing safety concerns regarding AVs, especially with regard to ``technical bugs'' or ``growing pains.''  It also shows that respondents are, generally, well aware of a variety of important social and ethical issues regarding the deployment of AV technology.

\subsection{Macro-Ethics: Society and Workforce}
\label{sec:Macro}

The UL drivers perceived that AV technology would be deployed in 10 years, and they had concerns regarding safety, social, and ethical issues.  Besides job losses, they were concerned that early deployment could create many other problems.  Respondents pointed out that some new technologies might work in experiments with controlled environments, but the results could be considerably different outside those limits.  Pedestrians, for instance, were seen as being negatively affected by automation; drivers stated that human intuitions were better suited to pedestrian interactions than AVs.  As one of the UL drivers stated, ``People have run in front of my car while I'm having a conversation and it's hard to balance that.  [\ldots] They might not really take that [a car is autonomous] into account and that could lead to accidents'' (UL-1).  This driver raised his concerns regarding pedestrians and industry over-reach emphatically: ``There's no coincidence that Google Waymo is testing their technology in Scottsdale.  The city planning there is very grid-like and it's also very hot, during the day, which is the only time that Google Waymo is active, so there's less pedestrians, and less bicyclists out there which is another danger, so in places that have a grid plan, [it is easier to implement] [\ldots], but places like Boston it'll be [difficult].  Because there's a lot of city streets and curves and stuff like that [\ldots].  But, that doesn't mean companies won't try it'' (UL-1).

The intuition to protect others and react instinctively to avoid accidents is not the only human characteristic that UL drivers felt automation could not replicate.  They believed social practices would also be affected.  Another UL driver noted that ``most people do want to talk [\ldots even] with a random stranger [\ldots].  For them [the passengers] it's definitely probably going to be a little more soul-crushing to get into the car and see a box making the moves instead of you, or someone who can at least try to put on a friendly face'' (UL-3).  Other social changes include lack of agency.  As noted by this UL driver, ``I think we're going to get into a generation where\ldots it makes us lazy and\ldots dependent'' (UL-4).  Semi-professional drivers felt that the \emph{dehumanization of driving} would be a loss caused by AV technology.

The TR drivers expressed much greater skepticism about AVs.  They perceived that AVs may be deployed in 10 to 20 years and identified a range of concerns.  Some of these concerns were similar to those of the UL drivers, yet their concerns were often more nuanced.  As one TR driver stated, ``I don't know as if the driver will ever be fully replaced.  [\ldots] I think for interstate travel where it's highways and open roads, that's gonna happen first but when you're in the city and suburbs and there's more variability [\ldots] on an interstate highway you don't have pedestrians that you have to worry about'' (TR-1).

Some of the TR drivers were vehemently opposed to the AV technology.  TR driver Christina was highly critical, declaring that: ``I think it [AV Tech] would be wrong, and it would put a lot of people in danger'' (TR-3).  There is some reluctance among TR drivers critical toward AVs to even estimate when the technology might be widely adopted.  In her words, ``I hope never.  I don't know if it will be.  I hope not soon'' (TR-3).  Her criticism of automation is specific and based on concrete experiences with automation technology: ``The robots load you.  The robots put the things in your trailer.  Everything was done by robots.  Until delivery when you find out they didn't put the slab right or they forgot something else.  [\ldots] before you used to be able to go inside and look through the load to have it how you like it to be loaded, now the machine goes in and loads you'' (TR-3).  The drivers seem to base their negative opinion of AV technology on personal experiences with automation and highly publicized tragedies involving self-driving technology: ``What is going to happen when [\ldots] you let that computer drive you?  [\ldots] A Tesla was in an accident and the car completely shut down; you couldn't open the window.  [\ldots] He burned completely.  The car was on fire and the fire department couldn't take him out, because it was all electrical and it all shut down.  So, I think that is completely wrong'' (TR-3). 

Other professional drivers offered a more mixed assessment with detailed or nuanced estimates.  TR driver Bill estimated that ``the first 10 years are gonna be driver and technology both together\ldots then I'm sure at one point we will see it to where it will completely take over trucking'' (TR-2).  Some drivers even provided state specific estimates: ``I'd say 10-15 years maybe, in the right places, in the right states.  Like I'd see them driving around in Texas or Nebraska, but Colorado or Wyoming where there's white out conditions?  I don't see them dealing with that'' (TR-4).

An overarching concern was that \emph{their voices would not be heard at any stage of public policy deployment}.  As one TR driver noted, ``When you have people sitting there making rules like that, who have never been in the truck---it's just mind boggling how they get away with this?''  (TR-4).

\subsection{Evaluating the Initial Claims}
\label{sec:Evaluating}

The results of our interviews help assess the claims introduced in Section~\ref{sec:Claims}.  Does our interview data confirm, disconfirm, or neither confirm nor disconfirm the relevant claim?  In sum, the claims about the \fsl{transportation industry} and \fsl{drivers' expectations} are confirmed; the claims about \fsl{reskilling} and \fsl{responsibility to respond} are neither confirmed nor disconfirmed; the claim about \fsl{driving as a profession} is disconfirmed.

\begin{description}
\item[Transportation industry] \emph{AVs significant impact on transportation is inevitable: \fbf{Confirmed}}.  For TR respondents, this is confirmed by all except Christina (TR-3), who hopes it does not ever come.  UL respondents agree wholeheartedly that AVs will inevitably impact the way in which the transportation system is managed.

\item[Drivers' expectations] \emph{Employers should be straightforward about changes and options toward the potentially affected stakeholders: \fbf{Confirmed}}.  For TR respondents, this is the case across the board.  They referenced the electronic logbook regulations that had recently been put into place, with which they were frustrated, and they all spoke of the importance of communication in regard to changes in the industry and associated regulations.  UL drivers unanimously agreed as well, although it was less of a concern for them than for TR drivers.  Two UL drivers were interested in the recent California law that established ``gig-economy workers'' as employees, rather than as contractors.  The two respondents, UL-1, and UL-4, were on opposite sides of opinion regarding the law, but both agreed that greater consideration for workers was necessary.  UL-4 expressed outright distaste for the communication (or lack thereof) received from Uber about new changes to Uber's policy.  UL-1 expressed contempt for Uber and Lyft, calling them both ``slimy'' companies.  The other UL drivers all anticipated changes coming and expressed their approval for better communication with drivers.

\item[Reskilling] \emph{Reskilling is manageable due to the decade-plus lead time: \fbf{Neither}}.  This claim is marginally correct.  Half of the TR respondents agreed with this claim, whereas two others (TR-3 and TR-4) were either dismissive of all automation, or frustrated that so many jobs would be lost, respectively.  The UL drivers viewed the advancement of AVs more positively, and were less likely to care about reskilling, as they used the income merely to supplement their livelihood.  To the UL drivers, it was seen mostly as a necessary, but unfortunate, step into the future.

\item[Responsibility to respond] \emph{Responsibility for dealing with these issues falls across all elements of society (government, business, drivers): \fbf{Neither}}.  The discussion of responsibility varied across respondents, and no clear trend was observed.

\item[Driving as a profession] \emph{To the extent that driving jobs are ``unsatisfying and potentially unhealthy'' their elimination could be a positive development so long as other opportunities are available: \fbf{Disconfirmed}}.  Not one driver from either category said this was the case.  Each said that driving for them was a net positive, and that they enjoyed what they did.  Some interesting caveats were brought forth by UL-1.  This respondent gives stipulations, such as being a college student, who is comfortable defending themselves, but says that driving for Uber as a full-time job and only source of income would be an ``unimaginable hell.''  However, every professional driver found their job to be positive, despite the challenges, even going as far to say that technology and the elimination of their jobs would, in fact, be a negative.  Overall, our interviewee data disconfirms the idea that drivers found their jobs unsatisfying or unhealthy.

\end{description}

\section{Conclusions and Directions}
\label{sec:Conclusions}

We claimed that stakeholders' perspectives constitute important data in a full assessment of the social and ethical impact of AVs.  Toward that end, we engaged with professional and semi-professional drivers using a semi-structured interview methodology.  These drivers are aware of and concerned with both micro and macro-ethical issues in relation to AVs, both of which motivate this investigation and which society should actively address.  By considering drivers, we obtained results that disconfirm some of the findings of studies of experts.

We view the interviewee data analyzed in this paper as informing the thought process of parties in Rawls' original position, particular in later stages as specific social and economic policies are developed.  The perspectives and opinions of drivers (and stakeholders in other relevant fields impacted by AI) are clearly morally relevant to the kinds of principles that should be devised in the original position.  To be clear, we are not suggesting that our drivers were, or should be, placed behind a veil of ignorance during interviews.

Instead, we suggest that drivers' lived experience and perspectives should be part of a general basis of knowledge about the effects of AVs on transportation and the economy, because this better informs the design of principles and polices to govern deployments of AV technology.  On the most practical level, decision makers should pay close attention to these perspectives as they develop policies.  A reasonable principle of fairness requires it.

Additional studies that focus on diverse stakeholder groups, such as pedestrians, public transport users, and people with mobility limitations, are necessary to fully align AV technology's deployment with values of the public at large.  Doing so can improve the ethical standing of our policies and increase public confidence that values of the relevant stakeholders are being respected.

Our findings raise interesting challenges for further investigation.  Along the theoretical dimension is advancing theories of justice and fairness in light of the permeation of AI into society.  Along the practical dimension are identifying the elements of work processes that would be lost due to automation.  The future is promising for AI and for deeper analysis of the ethics of AI from the perspectives of Science and Technology Studies.




\end{document}